\documentclass[12pt]{iopart}

\usepackage{graphicx}
\usepackage{iopams}
\usepackage[numbers, sort&compress]{natbib}

\begin{document}

\title [Percolation fronts with spatially varying densities]
{The geometry of percolation fronts in two-dimensional lattices with spatially
  varying densities}

\author{Michael T Gastner$^{1,2}$ and Be\'ata Oborny$^3$}
\address{$^1$Department of Mathematics, Complexity and Networks Programme,
  Imperial College London, South Kensington Campus, London SW7 2AZ, United
  Kingdom}
\address{$^2$Department of Engineering Mathematics, University of
  Bristol, Merchant Venturers Building, Woodland Road, Bristol BS8
  1UB, United Kingdom.}
\address{$^3$Department of Plant Taxonomy, Ecology and Theoretical
  Biology, E\"otv\"os Lorand University, P\'azm\'any s\'et\'any 1/C,
  Budapest, H-1117, Hungary}
\ead{m.gastner@imperial.ac.uk}

\begin{abstract}
  Percolation theory is usually applied to lattices with a uniform
  probability $p$ that a site is occupied or that a bond is closed.
  The more general case, where $p$ is a function of the position $x$,
  has received less attention.
  Previous studies with long-range spatial variations in $p(x)$ have
  only investigated cases where $p$ has a finite, non-zero gradient at
  the critical point $p_c$.
  Here we extend the theory to two-dimensional cases in which the
  gradient can change from zero to infinity.
  We present scaling laws for the width and length of the hull
  (i.e.\ the boundary of the spanning cluster). 
  We show that the scaling exponents for the width and the length
  depend on the shape of $p(x)$, but they always have a constant ratio
  $4/3$ so that the hull's fractal dimension $D=7/4$ is invariant. 
  On this basis, we derive and verify numerically an asymptotic expression for
  the probability $h(x)$ that a site at a given distance $x$ from $p_c$
  is on the hull.
\end{abstract}
\pacs{05.50.+q, 05.45.Df, 05.70.Jk}
\maketitle	

\section{Introduction}

\begin{figure}
  \begin{center}
    \includegraphics[width=8.6cm]{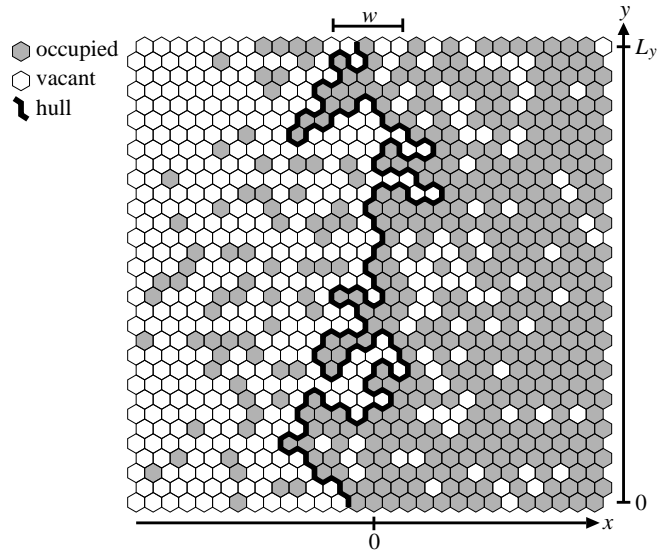}
    \caption{
      Site percolation on a triangular lattice where occupied
      sites (grey hexagons) are more abundant towards increasing
      $x$-coordinates. 
      The hull (black curve) marks the boundary between the vacant
      spanning cluster on the left and the occupied spanning cluster on
      the right.
      The hull width $w$, defined in \eref{width}, is the standard
      deviation in the hull's $x$-coordinates.
    }
    \label{triperc}
  \end{center}
\end{figure}

Percolation theory is a powerful tool in describing systems of randomly
located objects that are placed on the sites or bonds of a
network~\cite{StaufferAharony91}.
The sites or bonds are independently occupied with a probability $p$
or remain vacant with probability $1-p$.
The task of percolation theory is to identify the properties of the
clusters formed by the connected components.
In most theoretical studies, $p$ is assumed to be equal everywhere
(i.e.\ the system is assumed to be uniform).
However, in many real-world applications, it is more natural to allow some
variation in $p$ among the sites or bonds.
In spatial systems, $p$ often depends on the position.
For example, if occupied sites represent particles that diffuse from the
right-hand edge of a two-dimensional lattice and are absorbed on the left-hand
edge, then the concentration of particles exhibits a gradient~\cite{Sapoval_etal85}
and the system looks qualitatively like the lattice in figure \ref{triperc}.
Gradients also frequently arise in the spatial distribution of
biological populations because the environmental conditions 
may change gradually in space (e.g.\ from lower to higher elevation
along a hillside).
If occupied sites represent the presence of a species, the population
can exhibit a spatial transition from isolated individuals to a large
connected component~\cite{Milne_etal96,Gastner_etal09}.

\begin{figure}
  \begin{center}
    \includegraphics[width=8.6cm]{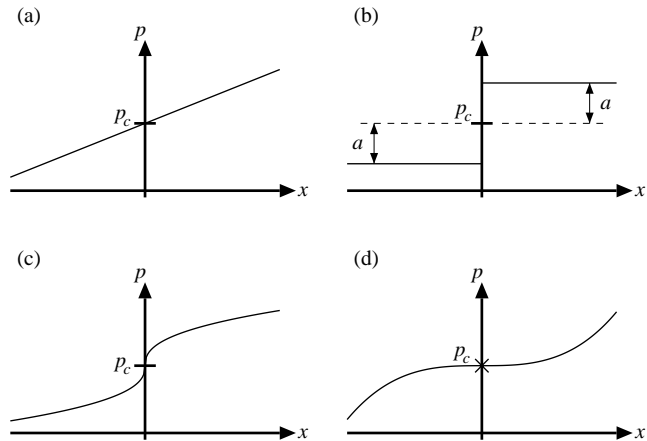}
    \caption{
      The studied occupancy probabilities $p(x)$. 
      (a): Linear function. This has been, at least as a local
      approximation, the assumption in all previous studies of
      gradient percolation.
      (b): Discontinuous function ($b=0$ in \eref{p_power}).
      (c) and (d): Continuous function with infinite or zero slope at $x=0$
      ($0<b<1$ or $b>1$, respectively).
    }
    \label{p_vs_x}
  \end{center}
\end{figure}

Given such a transition in connectivity, the focus of gradient
percolation is on the hull~\cite{Voss84}, defined as the interface
between adjacent
occupied and vacant clusters that both fully span the lattice's
$y$-direction (black curve in figure \ref{triperc}).\footnote{If the
  length $L_y$ of the lattice in the $y$-direction is finite, then
  there is a finite probability to find more than one pair of adjacent occupied
  and vacant spanning clusters and hence multiple hulls.
  However, this probability goes rapidly to zero as
  $L_y\to\infty$~\cite{Grassberger02}, see \cite{Nolin08_AOP} for a
  rigorous proof.
  We never encountered this situation in our numerical simulations
  with $L_y\geq 4096$.
} 
The hull has been used in practice, for example, in studies of the
surface structure of polymers~\cite{WoolLong93}, of the treeline
dividing woodland from grassland~\cite{Milne_etal96,Gastner_etal09},
of the boundary of urban settlements~\cite{Makse_etal98} and as an
example of a non-trapping self-avoiding random
walk whose properties have been investigated
theoretically~\cite{WeinribTrugman85,SaleurDuplantier87} and with computer simulations~\cite{Ziff_etal84,Ziff86}.
Apart from its many applications, gradient percolation is also an
active field of theoretical investigation.
It allows highly accurate measurements of the smallest occupancy
probability $p_c$ for which an infinite spanning cluster of occupied sites exists.
The value of $p_c$ is equal for gradient and non-gradient (i.e.\ uniform)
lattices.
Introducing a gradient has the advantage that $p$ does not need to be
carefully fine-tuned to achieve excellent numerical
results~\cite{Rosso_etal85,ZiffSapoval86}.
Furthermore, the geometry of the hull is particularly intriguing.
Suppose the hull connects the sequence of coordinates
$(x_1,y_1=0),\ldots,(x_l,y_l=L_y)$ delineating the boundary of the
spanning cluster, where $L_y$ is the system size in the $y$-direction.
If $p$ can be approximated near $p_c$ by a linear function
$p(x)=ax+p_c$, then the hull width $w$, defined as the standard
deviation of the hull's $x$-coordinates,
\begin{equation}
  w=\sqrt{\frac{\sum_i x_i^2}l - \left(\frac{\sum_i x_i}{l}\right)^2},
  \label{width}
\end{equation}
is known to scale as $a^{-4/7}$ in the limit $a\to 0$.
The hull length $l$ is proportional to $L_y a^{-3/7}$ and, on length
scales below $w$, the hull's fractal dimension is
$D=7/4$~\cite{Sapoval_etal85}.
These non-trivial scaling laws
-- recently proved rigorously by Nolin
\cite{Nolin08_AOP, Nolin08_EJOP} --
 have sparked a lot of interest (see
\cite{GouyetRosso05} for a survey of the literature)
and some researchers have generalized the results to gradients in more complicated
lattice models~\cite{Kolb_etal87,BoissinHerrmann91,Lemarchand_etal96,Sapoval_etal04,Loscar_etal09,Gastner_etal11}.
So far, all studies have assumed that $p(x)$ can be treated, at
least locally, as a linear function (figure~\ref{p_vs_x}(a)).
However, this is frequently not the case in reality.
As $a\to 0$, higher order terms in the Taylor expansion of $p$ may
become dominant so that $w$ and $l$ do not diverge, although this is
predicted when interpreting $w\propto a^{-4/7}$ and $l\propto
L_ya^{-3/7}$ literally.
Local linearization is not possible either if the gradient does not
exist at all, for instance if $p$ jumps discontinuously from one value to another.

Here we examine representative cases in which no linear approximation
for $p(x)$ exists at $p_c$. 
We investigate the class of functions
\begin{equation}
  p(x) =
  \cases{
    \max(0,\,p_c - a(-x)^b)&if $x<0$,\cr
    \min(1,\,p_c + ax^b)&if $x\geq 0$\cr}
  \label{p_power}
\end{equation}
for $b\geq 0$ and $a>0$. 
If $b=0$, $p(x)$ makes a discontinuous jump over the percolation
threshold $p_c$ (figure \ref{p_vs_x}(b)).\footnote{
  For functions defined only on the discrete coordinates of a lattice,
  we cannot, strictly speaking, distinguish continuous from
  discontinuous behaviour.
  However, for $p(x)$ defined in \eref{p_power}, we can apply the
  following criterion.
  Let $x_-$ and $x_+$ be the neighbouring left and right coordinates
  of $x$ in the lattice.
  If and only if $p(x)\to p(x_-)$ and $p(x)\to p(x_+)$ in the limit $a\to0^+$, then
  the underlying function $p$ of real-valued coordinates must be continuous at $x$.
  In this case, we also call the discretized function continuous.
}
For $0<b<1$, $p(x)$ is continuous, but the gradient at $p_c$ is
infinite (figure \ref{p_vs_x}(c)).
If $b>1$, the gradient is zero (figure \ref{p_vs_x}(d)), so that
$|p(x)-p_c|$ approaches zero faster than any linear function at $x=0$.
We have chosen the coefficient $a$ in \eref{p_power} to be equal on
the positive and negative $x$-axis.
This is a convenient choice because, as we show below, the
probability to find the hull at a coordinate $x$ is in this case
symmetric about $x=0$.
The results described below can be easily generalized to cases where
the coefficients differ in the positive and negative direction. 
We present numerical simulations for site percolation on triangular
lattices, where $p_c=1/2$, and apply periodic boundary conditions in
the $y$-direction ($L_y\geq 4096$).
For a comparison, we also performed simulations for site and bond
percolation on square lattices.
We found the same scaling exponents, indicating that they are
independent of the lattice 
type and presumably also valid in continuum models~\cite{Quintanilla_etal00}.

\section{Hull width, length and fractal dimension}

\begin{figure}
  \begin{center}
    \includegraphics[width=5cm]{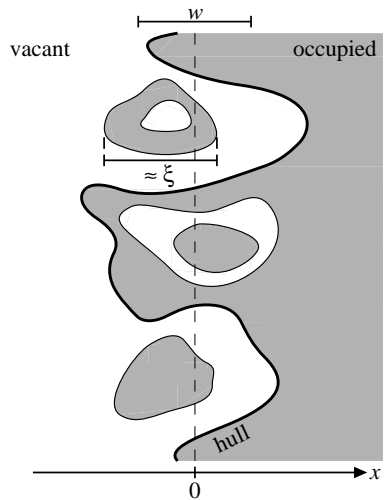}
    \caption{
      Illustration of the relation between hull width $w$ and correlation
      length $\xi$ near $x=0$, where the hull meanders between islands
      of vacant and occupied sites.
      The linear size of these obstacles is of the order of $\xi$.
      Therefore, the hull width $w$, which measures the fluctuations
      of the hull around $x=0$, is also approximately $\xi$.
    }
    \label{wdth_scal_explan}
  \end{center}
\end{figure}

\begin{figure}
  \begin{center}
    \includegraphics[width=8.6cm]{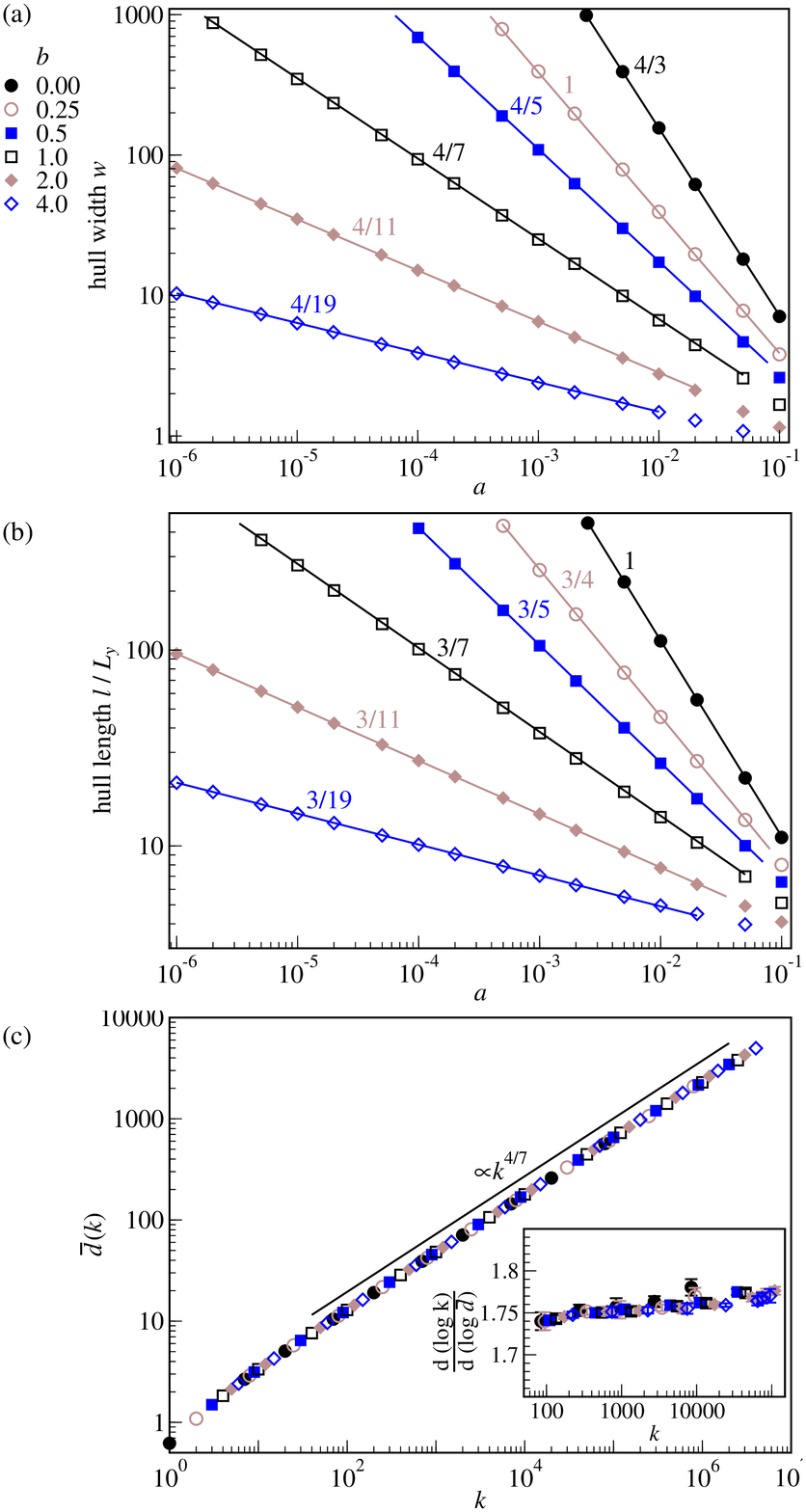}
    \caption{
      (a) The hull width $w$ versus $a$ for different values of $b$ in
      \eref{p_power} from computer simulations (shown as circles,
      squares, and diamonds). 
      The lines indicate the expected power law $w\propto a^{-B}$ with $B=\nu/(\nu
      b+1)$. 
      The number next to each line states the value of $B$.
      Error bars are smaller than the symbol sizes.
      (b) The normalized hull length $l/L_y$ scales $\propto a^{-C}$ with
      $C=1/(\nu b+1)$ as displayed by the lines.
      (c) The fractal dimension $D$ of the hull is determined with the
      equipaced polygon method~\cite{KayeClark85} described in
      the appendix.
      In this method, we compute the mean Euclidean distance
      $\bar{d}(k)$ between hull coordinates that are $k$ steps apart.
      A numerical estimate of $D$ can be obtained by fitting
      $\bar{d}(k)\propto k^{1/D}$.
      We observe that $D=7/4$ is independent of $b$.
      Inset: $D=1.75$, numerically calculated as the derivative $\rmd
      (\log k)/\rmd (\log\bar{d})$, is within the error bars for
      almost three orders of magnitude of $k$.
    }
    \label{wdth_lngth_df}
  \end{center}
\end{figure}

Let us first consider the hull width $w$ in the discontinuous case ($b=0$ in
\eref{p_power}).
Because $a>0$, the hull must navigate around those occupied clusters to the left
that are not part of the occupied spanning cluster (figure \ref{wdth_scal_explan}).
Similarly, on the right-hand side, excursions of the hull must steer
clear of smaller vacant islands that have not merged with the vacant spanning cluster.
The typical linear size of these smaller clusters is given by the
correlation length $\xi$ which scales in uniform systems
$\propto|p-p_c|^{-\nu}$ if $p\approx p_c$.
Because the critical exponent $\nu$ is equal to $4/3$ in two
dimensions~\cite{StaufferAharony91,SmirnovWerner01} and
because the hull width scales in proportion to the cluster sizes, we
expect $w\propto a^{-4/3}$.
Our numerical simulations confirm this intuition (black filled circles in
figure \ref{wdth_lngth_df}a).

If $b>0$, the situation is more complicated because $p(x)$, and hence
also $\xi(p(x))$, change gradually as a function of $x$.
For $b=1$, Sapoval et al.\ \cite{Sapoval_etal85} have pointed out that
we can still derive the correct scaling behaviour by assuming that $w$
is proportional to the correlation length at $w$.  
The self-consistent relation $w \propto \xi(p(w))$, where the local
correlation length satisfies $\xi(p(x))\propto |p(x)-p_c|^{-\nu}$
and $p(x)=p_c+ax$, yields the aforementioned $w\propto a^{-\nu/(\nu+1)} = a^{-4/7}$
(black open squares in figure \ref{wdth_lngth_df}a).
Carrying out the calculation more generally for $p(x)$ given by
\eref{p_power} with arbitrary $b$ and assuming $a$ to be small, we
obtain
\begin{equation}
  w\propto a^{-\nu/(\nu b+1)}.
  \label{general_w_scal}
\end{equation}
The same self-consistency requirement has been noticed and tested for
the Ising quantum chain~\cite{Platini_etal07,Collura_etal09}, where
the exponent $\nu$ however has a value different from the exponent in percolation.
In figure~\ref{wdth_lngth_df}(a), we compare data from numerical
simulations with the prediction of equation \eref{general_w_scal} based on
the two-dimensional percolation exponent $\nu=4/3$.
The data show excellent agreement. 

There is a similar scaling relation between the normalized hull length
$l/L_y$ and $a$, where $l$ is the number of steps in the hull in a
lattice of vertical dimension $L_y$.
The scaling of $l/L_y$ cannot be derived as intuitively as
the scaling of the hull width. 
However, previous numerical evidence in the case $b=1$ has supported
the hypothesis $l/L_y\propto a^{-3/7}$~\cite{Sapoval_etal85}, now
known to be correct~\cite{Nolin08_AOP}.
For arbitrary $b$, our data (figure \ref{wdth_lngth_df}b) indicate the
more general relation
\begin{equation}
l/L_y\propto a^{-1/(\nu b+1)},
\label{general_l_scal}
\end{equation}
which includes the scaling law for $b=1$ as a special case.

The scaling of $l$ and $w$ is related to the fractal dimension $D$ of
the hull as follows.
For a given value $a=a_1$ and system size $L_{y,1}$, let us denote
the hull length by $l_1$ and the width by $w_1$.
If we coarse-grain the hull by measuring it with rulers whose length is $A$
times longer than the lattice constant, we need $l_2\propto A^{-D}l_1$ 
rulers to cover the full length of the hull~\cite{Mandelbrot67}.
The width and the system size, however, are linear objects: it requires
$w_2\propto A^{-1}w_1$ rulers to span across the hull's $x$- and $L_{y,2}\propto
A^{-1}L_{y,1}$ rulers to cover its $y$-extent.
Thus, measured in units that are increased by a factor $A$, the hull
length, width and system size all appear smaller by either a factor
$A^{-D}$ or $A^{-1}$.
Now suppose that the hull geometry scales as $w\propto a^{-B}$
and $l\propto L_y a^{-C}$.
We obtain the same $w_2$ and $L_{y,2}$ as before if the lattice
constant remains our fundamental length scale, but we first replace
$a_1$ by $a_2\propto A^{1/B} a_1$ and then, from the original
system of length $L_{y,1}$, we cut out a strip of size $A^{-1}L_{y,1}$.
In this geometry $l_2\propto (A^{-1}L_{y,1})(A^{-C/B}a_1^{-C})\propto A^{-(1+C/B)}l_1$.
Comparing this with our previous expression for $l_2$, we conclude
that
\begin{equation}
  D = 1 + \frac{C}{B}.
  \label{fracdim}
\end{equation}
Inserting $B=\nu/(\nu b+1)$ and $C=1/(\nu b+1)$ from equations
\eref{general_w_scal} and \eref{general_l_scal} we find $D=1+1/\nu =
7/4$, independent of $b$.
We also measured the fractal dimension directly using an independent
numerical method (following \cite{KayeClark85,Ziff86}
and outlined in the appendix) and again find good
agreement with $D=7/4$ (figure \ref{wdth_lngth_df}c).
The fractal dimension of cluster hulls has recently received a lot of interest.
In uniform percolation, where $p=p_c$ at all sites (i.e.\ $a=0$ in
\eref{p_power}), $D=7/4$ is now known to be exact, thanks to
mathematical advances in stochastic Loewner
evolution~\cite{SmirnovWerner01}.
Recent work by Nolin and Werner \cite{NolinWerner09} proves that the
Hausdorff dimension remains $7/4$ if $p$ is near, but not exactly
equal to $p_c$.
In constant gradients ($b=1$ in \eref{p_power}), $D=7/4$ was
already conjectured in early studies~\cite{Sapoval_etal85} and related
to the scaling of $w$ and $l$.
Our results demonstrate that $D=7/4$ holds much more generally.
Although $B=\nu/(\nu b+1)$ and $C=1/(\nu b+1)$ depend on $b$ and thus
differ in the different scenarios depicted in figure \ref{p_vs_x}, the
ratio $B/C=\nu$ and hence the fractal dimension $D=1+1/\nu$ are
invariant properties of the hull geometry.

\section{Hull density profile}
\label{hull_density_profile}

\begin{figure}
  \begin{center}
    \includegraphics[width=8.6cm]{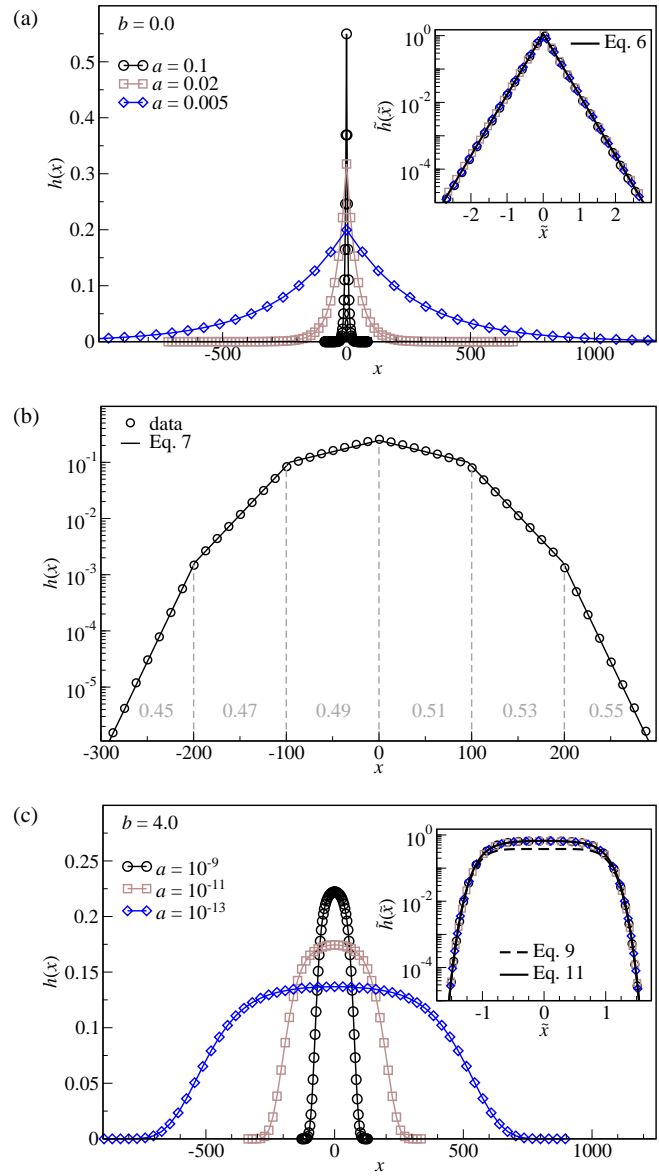}
    \caption{
      (a) Hull density $h$ as a function of position $x$ for
      $b=0$ in equation \eref{p_power}. Inset: data collapse for
      $\tilde{h}(\tilde{x}) = a^{(1-\nu)(\nu b+1)}h(a^{-\nu/(\nu
        b+1)}\tilde{x})$.
      (b) Hull density for a piecewise continuous function $p(x)$ with multiple
      discontinuities. The local value of $p(x)$ is given by the grey
      numbers under the graph. In every interval of constant $p$, $h$
      is approximately exponential, resulting in straight lines in
      this logarithmic plot.
      (c) Hull density for $b=4$.
    }
    \label{gen_norm}
  \end{center}
\end{figure}

The hull width and length are simple measures of the hull geometry,
but the distribution of the hull's $x$-coordinates contains yet more
information.
For fixed $x$, we define the hull density $h(x)$ as the fraction of
coordinates $(x,y)$ in the lattice that are on the hull.
We focus first on the case $b=0$ where $p(x)=p_c\mp a$ provided that $a$ is
sufficiently small.
The $\mp$ sign is to be interpreted as a minus (plus) sign for
negative (positive) $x$.
Our simulations for $b=0$ show that $h(x)$ has a maximum at $x=0$ and falls off
exponentially and symmetrically to both sides (figure \ref{gen_norm}a),
\begin{equation}
  h(x) = E_0 a^{\nu-1} \exp(-F a^\nu |x|),
  \label{exp_decay}
\end{equation}
where $E_0$ and $F$ are positive constants. 
The exponential decay of $h(x)$ is reminiscent of the off-critical
correlation function $g_c(\mathbf{r})\propto\exp(-|\mathbf{r}|/\xi)$ in uniform
percolation, where $g_c(\mathbf{r})$ denotes the probability that
$\mathbf{r}$ and the origin are part of the same finite cluster.
The decay rate $(Fa^\nu)^{-1}$ has the same scaling as the correlation
length $\xi$, namely $\propto |p-p_c|^{-\nu}$ for both $p<p_c$ and $p>p_c$.
However, there is one important difference between $(Fa^\nu)^{-1}$ and the
correlation length.
In uniform percolation, $\xi$ has different amplitudes on the sub- and
supercritical branch~\cite{Delfino_etal10}.
By contrast, we observe that $F$ is equal for $x<0$ and $x\geq 0$.

Equation \eref{exp_decay} has several interesting consequences.
First, $h(x)$ is a symmetric function (i.e.\ $h(x)=h(-x)$).
We observed this symmetry in all investigated lattice types
(triangular site percolation, square site and bond percolation).
Second, \eref{exp_decay} is consistent with the scaling relations
\eref{general_w_scal} and \eref{general_l_scal} of the width and
length for $b=0$.
This can be proved by inserting \eref{exp_decay} into
$l=\int_{-\infty}^\infty h(x)dx$ and $w^2=\int_{-\infty}^\infty
(x^2h(x)dx)/l$.
Third, after appropriate rescaling of $h$, the function
$\tilde{h}(\tilde{x}) = a^{1-\nu}h(a^{-\nu}\tilde{x})$ is independent
of $a$ (see the data collapse in the inset of figure \ref{gen_norm}(a)).

The case $b>0$, where $p(x)$ changes continuously, requires a little more
effort than $b=0$, where $p(x)$ is constant except for one discontinuity.
Our goal is to approximate $h(x)$ for $b>0$ by discretizing $p(x)$.
To this end, let us consider the intermediate case where $p$ has
more than one discontinuity: $p(x) = 0$ for $x<x_1$, $p(x) = \pi_1$ for
$x\in[x_1,x_2)$, $\ldots$, $p(x)=\pi_{n-1}$ for $x\in[x_{n-1},x_n)$
and $p(x) = 1 $ for $x_n\leq x$ with $0<\pi_1<\ldots\pi_{n-1}< 1$. 
Let us assume that the length $(x_i-x_{i-1})$ of each plateau is larger
than the exponential decay length $(F|\pi_i-p_c|^{\nu})^{-1}$. 
In this case, $h$ consists of piecewise exponential
functions $\propto \exp(\pm F |\pi_i-p_c|^\nu x)$ that are continuously
joined together at the breakpoints $x_1,\ldots,x_n$ (an example is
shown in figure \ref{gen_norm}b).
In terms of a differential equation, we can express this as
\begin{equation}
  \frac{\rmd}{\rmd x}\log h(x) = \pm F|p(x)-p_c|^\nu,
  \label{dlogh_dx_discrete}
\end{equation}
where $F$ is a constant.
The plus sign holds for $p(x)<p_c$; otherwise the minus sign applies.

We now make two approximations to obtain an analytic expression for
$h(x)$ if $p(x)$ changes continuously.
First we assume that the rate of change in $p(x)$ is small compared to
the characteristic decay length $(F|p(x)-p_c|^\nu)^{-1}$ for all $x$.
This allows us to insert \eref{p_power} directly into
\eref{dlogh_dx_discrete}. We obtain
\begin{equation}
\frac{\rmd}{\rmd x} \log h(x) = \pm F a^\nu (\mp x)^{\nu b}
\label{dlogh_dx_continuous}
\end{equation}
for $x$ in the interval $[-(p_c/a)^{1/b}, ((1-p_c)/a)^{1/b}]$.
We know from \eref{general_w_scal} that the hull width is a
vanishing fraction of the width of this interval for $a\to 0^+$.
Thus, our second approximation consists of extending \eref{dlogh_dx_continuous}
over the entire real line.
The solution is
\begin{equation}
  h(x) = E_b a^{(\nu-1)/(\nu b + 1)}\exp\left(-\frac{F}{\nu b+1} a^\nu |x|^{\nu b + 1}\right),
  \label{generalized_normal}
\end{equation}
where we have factored out $a^{(\nu-1)/(\nu b + 1)}$ from the integration
constant so that, according to \eref{general_l_scal}, the
remaining factor $E_b$ does not depend on $a$.

\section{Discussion}

Straightforward integration proves that \eref{generalized_normal} is
consistent with the scaling of the hull width in
\eref{general_w_scal}.
Because \eref{generalized_normal} also satisfies the scaling of the
length \eref{general_l_scal}, it implies the correct fractal dimension
$7/4$ according to \eref{fracdim}.
In addition to the scaling laws, \eref{generalized_normal} also provides an
analytic expression for the functional form of the hull density profile:
$h(x)$ is, in this approximation, proportional to a generalized normal
distribution~\cite{Nadarajah05}.
For $b=1/\nu$, we predict a Gaussian profile.
Another special case is $b=0$, where we retrieve the Laplace
distribution of \eref{exp_decay}.
For $0\leq b<1/\nu$ the distribution is leptokurtic (i.e.\ it has a narrower peak
and fatter tails than a Gaussian); for $b>1/\nu$ it is platykurtic (broader
peak, thinner tails).

\begin{figure}
  \begin{center}
    \includegraphics[width=8.6cm]{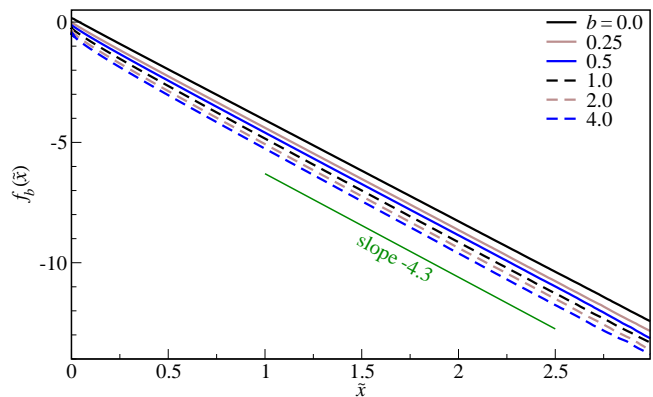}
    \caption{
      The transformation from $h$ to $f_b$ according to \eref{fb} shows that the
      hull density decays as $\exp(-Fa^\nu x^{\nu b+1}/(\nu b+1))$ with
      the same factor $F\approx 4.3$ for all $b$.
    }
    \label{super_scale}
  \end{center}
\end{figure}

For any particular $b$ in \eref{generalized_normal}, the rescaled
hull density $\tilde{h}(\tilde{x}) = a^{(1-\nu)/(\nu b +
  1)}h(a^{-\nu/(\nu b + 1)}\tilde{x})$ is independent of $a$ (insets
in figures \ref{gen_norm}(a) and \ref{gen_norm}(c)).
Rescaling can even go further. 
Above we have assumed that the constant $F$ in \eref{dlogh_dx_discrete} is
independent of $b$ so that the functions
\begin{equation}
  f_b(\tilde{x}) = \log\{a^{(1-\nu)/(\nu b+1)}h[((\nu
  b+1)a^{-\nu}\tilde{x})^{1/(\nu b+1)}] \}
  \label{fb}
\end{equation}
ought to be linear functions with the same slope $-F$ for all $b$.
Figure \ref{super_scale} demonstrates that this is indeed true for $\tilde{x}\gtrsim
1$, which is valid if $x\gtrsim w$.
From linear regression to $f_b$, we estimate
$F=4.3\pm0.1$.\footnote{$F$ depends on the lattice type, unlike $B$,
  $C$ and $D$. The stated value is for triangular site percolation. For square site
  percolation, we find $F=3.9\pm0.1$, for square bond percolation $F=5.4\pm0.1$. 
}

For $b\ll1$, a simple line in figure \ref{super_scale} fits the data very
well for all $\tilde{x}>0$ and this is presumably exact for $b=0$.
However, as $b$ becomes larger, there are visible departures
from the asymptotic line for $\tilde{x}\ll1$.
In this regime, $p(x)$ varies significantly over length scales
comparable to the correlation length, thus violating the assumption
behind \eref{dlogh_dx_continuous}.
If $b\gg 1$, we are effectively dealing with a system at the
critical point confined to the strip $-(p_c/a)^{1/b} <x< ((1-p_c)/a)^{1/b}$.
Some properties of critical cluster geometry in a strip are known
exactly~\cite{Schramm01,Simmons_etal07,IkhlefPonsaing12}.
From the known results for the ``one-pinch point
density'' in rectangular domains~\cite{Flores_etal12}, it is plausible that $h(x)$ is
analytic
and has a Taylor expansion around $x=0$ with leading terms $h(x) =
h(0)+h''(0)x^2/2+O(x^4)$ and $h''(0)<0$.
As a consequence, as $x\to0$ we expect $\log(h)$ to be approximately quadratic
(i.e.\ $\log(h(x))\approx G_0-H_0x^2$), in contrast to $\log(h(x))\approx
G_\infty-H_\infty |x|^{\nu b+1}$ for $|x|\to\infty$ as predicted by
\eref{generalized_normal}. 

\begin{figure}
  \begin{center}
    \includegraphics[width=8.6cm]{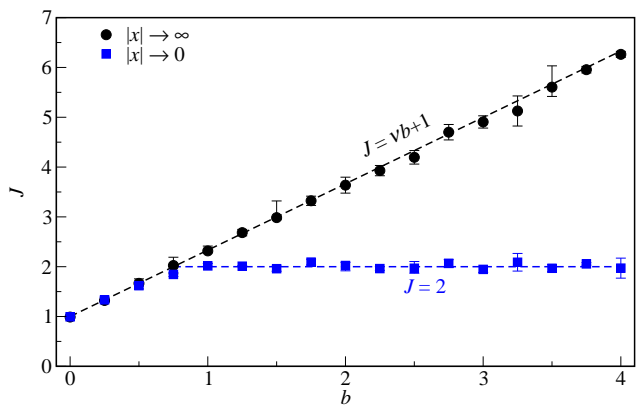}
    \caption{
      If the hull density is locally approximated as a function
      proportional to $\exp(-H |x|^J)$, the best fitting exponent
      $J$ increases as $\nu b+1$ in the tail (i.e.\ as $x\to\infty$), but is
      close to $2$ for $b>3/4$ in the core (i.e.\ as $x\to 0$).
    }
    \label{core_tail_exp}
  \end{center}
\end{figure}

We test this hypothesis with nonlinear least-squares regressions of
the general form $G-H|x|^J$ to fit either the core or the tail of $\log(h(x))$.
The best fitting exponents for $|x|\to\infty$ are indeed consistent with $J=\nu b+1$
(figure \ref{core_tail_exp}) for all $b$.
Performing the regression around $x=0$, on the other hand, we find a
crossover from $J=\nu b+1$ to $J=2$ at $b\approx1/\nu=3/4$.
This suggests that for $b<1/\nu$, \eref{generalized_normal} approximates
the hull density well, but for $b>1/\nu$ we should replace it with
\begin{equation}
  h(x) \approx E_ba^{(\nu-1)/(\nu b+1)}
  \exp\left(-\frac{F}{\nu b+1}a^\nu
    |x|^{\nu b+1} - K_b a^{2\nu/(\nu b+1)}x^2\right).
  \label{mixed_gauss_gennorm}
\end{equation}
This approximation has the observed asymptotic behaviour for both
$x=0$ and $|x|\to\infty$: $h(x)$ is Gaussian in the center, but drops more rapidly in
the tails.
In the inset of figure \ref{gen_norm}(c) we have used $F=4.3$ from above
and fitted $E_b$ and $K_b$ to the hull density for $b=4$.
The regression curve indeed fits the core of the distribution better than
\eref{generalized_normal} and also provides an excellent approximation in the tails.

\section{Conclusion}

In summary, we have investigated a generalization of gradient
percolation where the occupancy probability $p(x)$ is nonlinear at the
critical point $p_c$.
If, to lowest-order, $|p(x)-p_c| = a|x|^b$ for $a>0$ and $b\geq0$,
then the hull width and length scale as $w\propto a^{-\nu/(\nu b+1)}$
and $l\propto a^{-1/(\nu b+1)}$.
The hull's fractal dimension $D=7/4$ is independent of $b$.
The hull density is symmetric and drops for $x$ far away from $p_c$ in proportion to
$\exp(-Fa^\nu |x|^{\nu b+1}/(\nu b+1))$, where $F$ is a
lattice-dependent constant independent of $a$ and $b$.
Closer to $p_c$, the hull density becomes approximately Gaussian if $b>1/\nu$.

\ack
We thank G Pruessner and T S Evans for helpful comments.
MTG gratefully acknowledges support from Imperial College.

\section*{Appendix. Equipaced polygon method to determine the fractal dimension}
\label{eqpol}
There are many different ways to determine the fractal dimension of
the hull. 
We tested several of them and found that the equipaced polygon method
of \cite{KayeClark85} provides a good trade-off between simplicity and
consistent numerical results.
We begin by labeling all hull coordinates $(x_i,y_i)$ with a subscript
$i$ equal to the number of steps required to reach this position
along a walk on the hull starting at the bottom of the lattice.
At this stage, the hull is described by a sequence of coordinates
$(x_1,y_1=0)\ldots (x_l,y_l=L_y)$, where $l$ is the hull length.
For a given $k$ and for $1\leq i\leq l-k$, we then calculate the
Euclidean distances $d_i(k)$ between $(x_i,y_i)$ and $(x_{i+k},y_{i+k})$.
Finally we compute the mean $\bar{d}(k)=\sum_i d_i(k)/(l-k+1)$.
If the hull is a fractal of dimension $D$, we expect
$\bar{d}(k)\propto k^{1/D}$~\cite{Gastner_etal09}.

\end{document}